\documentclass[twoside,leqno,twocolumn]{article}

\usepackage[letterpaper]{geometry}

\usepackage{graphicx}
\usepackage{subcaption}
\usepackage{enumitem}
\usepackage{color}
\usepackage{amsmath}
\usepackage{ltexpprt}
\usepackage{algorithmic}

\newcommand{\qmarks}[1]{``#1''}

\begin{document}

\title{\Large Algorithms and Experiments Comparing Two Hierarchical Drawing Frameworks}

\author{Panagiotis Lionakis\thanks{Computer Science Department, University of Crete, GREECE.}
\and Giorgos Kritikakis\footnotemark[1]
\and Ioannis G. Tollis\footnotemark[1]}

\date{}

\maketitle

\begin{abstract} \small\baselineskip=9pt   
We present algorithms that extend the path-based hierarchical drawing framework and give experimental results.   Our algorithms run in $O(km)$ time,  where $k$ is the
number of paths and $m$ is the number of edges of the graph, and provide better upper bounds than the original path based framework: e.g., the height of the resulting drawings is equal to the length of the longest path of $G$, instead of $n-1$, where $n$ is the number of nodes.  Additionally, we extend this framework, by bundling and drawing all the edges of the DAG in $O(m + n \log n)$ time, using minimum extra width per path.
We  also  provide  some  comparison to a well  known  hierarchical  drawing  framework, widely known as the Sugiyama framework,  as a proof of concept.  The experimental results show that our  algorithms produce  drawings  that  are  better  in  area and number of bends, but worse for crossings in sparse graphs. Hence, our technique offers an interesting alternative for drawing hierarchical graphs. Finally, we present an $O(m + k \log k)$ time algorithm that computes a specific order of the paths in order to reduce the total edge length and number of crossings and bends.

\end{abstract}

\section{Introduction}
Hierarchical graphs are very important for many applications in several areas of research and business because they often represent hierarchical relationships between objects in a structure. They are directed (often acyclic) graphs and their visualization has received significant attention recently~\cite{DBLP:books/ph/BattistaETT99,kw,handbook}. An experimental study of four algorithms specifically designed for DAGs was presented in~\cite{DBLP:conf/gd/BattistaGLPTTVV96}. DAGs are usually used to describe processes containing some long paths, such as in PERT applications see for example~\cite{di1989automatic,fisher1983stochastic}. The paths can be either application based, e.g. critical paths or user defined. If one desires automatically generated paths, there are several algorithms that compute a path decomposition of minimum cardinality~\cite{DBLP:journals/siamcomp/HopcroftK73,DBLP:conf/recomb/KuosmanenPGCTM18,DBLP:conf/stoc/Orlin13,DBLP:journals/siamcomp/Schnorr78}. 
A new framework  to visualize directed graphs and their hierarchies is introduced in~\cite{ortali2018algorithms,JGAA-502}. It computes readable hierarchical visualizations in two phases  by \qmarks{hiding} (\emph{abstracting}) some selected edges while maintaining the complete reachability information of a graph. 
\par
In this paper we present polynomial time algorithms that follow the main framework of~\cite{JGAA-502} which is based on the idea of partitioning the vertices of a graph into \emph{paths/channels}, drawing the vertices in each path vertically aligned on some $x$-coordinate and then drawing the edges between vertices that belong to different paths. The produced drawings contain all edges of the input graph and attempt to optimize the height, width and number of bends of the resulting drawing. 
\par
This new framework departs from the typical Sugiyama framework~\cite{DBLP:journals/tsmc/SugiyamaTT81} and it consists of two phases: (a) Cycle Removal, (b) the path/channel decomposition and hierarchical drawing step. The Sugiyama framework has been extensively used in practice, as manifested by the fact that various systems are using it to implement hierarchical drawing techniques. Several systems such as \emph{AGD} \cite{DBLP:journals/spe/PaulischT90}, \emph{da Vinci} \cite{DBLP:conf/gd/FrohlichW94}, \emph{GraphViz} \cite{DBLP:journals/spe/GansnerN00}, \emph{Graphlet} \cite{DBLP:journals/spe/Himsolt00}, \emph{dot} \cite{gansner2006drawing}, \emph{OGDF}~\cite{DBLP:reference/crc/ChimaniGJKKM13}, and others implement this framework in order to draw directed graphs.  Commercial software such as the Tom Sawyer Software TS Perspectives \cite{Tom} and yWorks \cite{yWorks} essentially use this framework in order to offer automatic visualizations of directed graphs.  The comparative study of~\cite{DBLP:conf/gd/BattistaGLPTTVV96} concluded that the Sugiyama-style algorithms performed better in most of the metrics.  For more recent information regarding this framework see~\cite{handbook}.

Even though it is very popular, the Sugiyama framework has several limitations: as discussed bellow, most problems and subproblems that are used to optimize the results in various steps of each phase have turned out to be NP-hard. Additionally, several of the heuristics employed to solve these problems give results that are not bound by any approximation.  Furthermore, the required manipulations in the graph often increase substantially its complexity, e.g., up to $O(nm)$ dummy vertices may be inserted in a directed graph $G=(V,E)$ with $n$ vertices and $m$ edges. 
The overall time complexity of this framework (depending upon implementation) can be as high as $O((nm)^2)$, or even higher if one chooses algorithms that require exponential time.  Finally, another important limitation of this framework is the fact that heuristic solutions and decisions that are made during previous phases (e.g., crossing reduction) will influence severely the results obtained in later phases. Nevertheless, previous decisions cannot be changed in order to obtain better results. By contrast, in the main framework of~\cite{JGAA-502} most problems of the second phase can be solved in polynomial time. If a path decomposition contains $k$ paths, the number of bends introduced is at most $O(kn)$ and the required area is at most $O(kn)$. In order to minimize the number of crossings between cross edges and path edges the authors suggest checking all possible $k!$ permutations of the $k$ paths which may be reasonable for small values of $k$~\cite{ortali2018algorithms}.  However, edges between non consecutive vertices in a path, called \emph{path transitive edges} are not drawn in this framework. 
\par
Additionally, we offer experimental results comparing them to the results obtained by running the hierarchical drawing module of OGDF~\cite{DBLP:reference/crc/ChimaniGJKKM13}, which is based on the Sugiyama framework, and is the most updated research software that implements this framework.  Since the "cycle removal phase" is required in both frameworks,  we focus our experiments on the case where the input graph $G$ is acyclic (DAG). Our algorithms run in $O(km)$ time, and provide better upper bounds than the ones given in~\cite{JGAA-502}: (a) the height of the resulting drawings is equal to the length of the longest path of $G$, which is often significantly lower than $n-1$. (b) The \emph{path transitive edges} are drawn by our algorithms in such a way that the required extra number of columns is minimized for each path (see Section 3). The experimental results show that the drawings produced by our algorithms have a significantly lower number of bends and are much smaller in area than the ones produced by OGDF (see Section 4).  On the other hand, the drawings of OGDF have a lower number of crossings when the input graphs are relatively sparse. However, when the graphs are a bit denser (e.g., average degree greater than five) our drawings have less crossings.  Of course, it is expected that OGDF would be better than our algorithms in the number of crossings since OGDF places a significant weight in minimizing crossings, whereas we do not explicitly minimize crossings.  Thus our algorithms offer an interesting alternative to visualize hierarchical graphs.
Finally, we present an $O(m + k \log k)$ time algorithm that computes a specific order of the paths that further reduces the total edge length, and number of crossings and bends in sparse DAGs.

\section{Overview of the Two Frameworks}
\label{se:overview}

\begin{figure*}[!ht]
     \begin{subfigure}[l]{1.0\columnwidth}
         \centering
         \includegraphics[height=10cm ,width=0.6\linewidth]{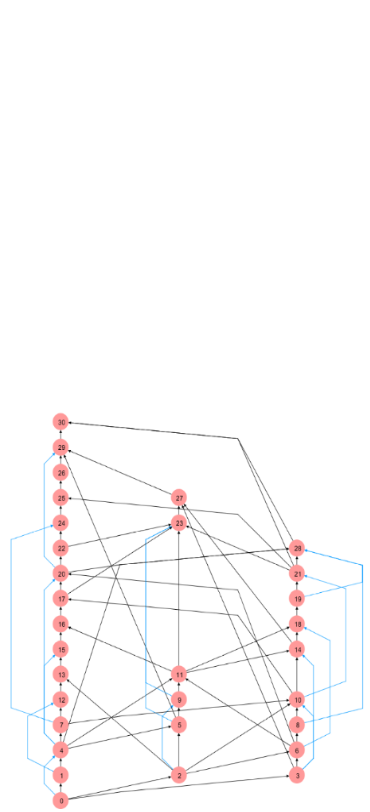}
           \caption{}
			\label{fig:sfig1}
     \end{subfigure}
     \hfill{}
     \begin{subfigure}[r]{1.0\columnwidth}
         \centering
         \includegraphics[height=10cm ,width=0.5\linewidth]{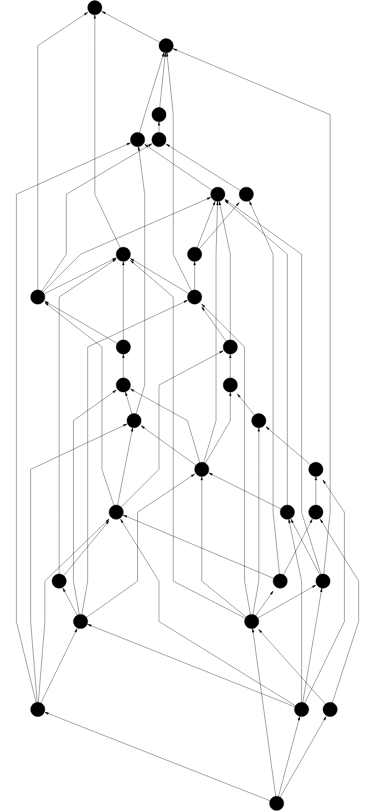}
           \caption{}
			\label{fig:sfig2}
     \end{subfigure}
	 \caption{In (a) we show the drawing $\Gamma$ based on $G$  as computed by Tom Sawyer Perspectives which follows our proposed framework. In (b) we show the drawing of the graph $G$ as computed by OGDF.}
\label{teaser}
\end{figure*}

In order to motivate our discussion about the two frameworks considered in this paper we present Figure~\ref{teaser} that shows a DAG $G$ drawn by these two frameworks:
Part~(a) shows a drawing $\Gamma$ of $G$ computed by our algorithms that customize the path-based framework of~\cite{JGAA-502}; it is implemented in Tom Sawyer Perspectives~\cite{Tom} (a tool of Tom Sawyer Software); part~(b) shows the drawing of $G$ computed by OGDF. The graph consists of 31 nodes and 69 edges. The drawing computed by our algorithms has 74 crossings, 33 bends, width 14, height 16, and area 224. On the other hand, OGDF computes a drawing that has 72 crossings, 64 bends, width 42, height 16 and area 672. The width and height reported by OGDF are 961 and 2273, respectively. We had to normalize these figures in order to have a reasonable comparison, as will be discussed later. 
As can be observed by these two drawings, the two frameworks produce vastly different drawings with their own advantages and disadvantages.


The Path Based Hierarchical Drawing Framework, call it \textbf{Algorithm PBH}, follows an approach to visualize directed acyclic graphs that \qmarks{hides} some edges and focuses on maintaining their reachability information~\cite{JGAA-502}. This framework is based on the idea of partitioning the vertices of the graph $G$ into (a minimum number of) \emph{channels/paths}, that we call \emph{channel/path decomposition} of $G$, which can be computed in polynomial time.   Therefore, it is orthogonal to the Sugiyama framework in the sense that it is a vertical decomposition of $G$ into (vertical) paths/channels and it consists of only two steps: (a) the cycle removal step (if the directed graph contains cycles) and (b) the channel decomposition and hierarchical drawing step. Thus, most resulting problems are \emph{vertically contained}, which makes them simpler, and reduces their time complexity. This framework does not introduce any dummy vertices and keeps the vertices of a path \emph{vertically aligned}.  By contrast, the Sugiyama framework performs a horizontal decomposition of a graph, even though the final result is a vertical (hierarchical) visualization. 

Let  $S_p= \{P_1,...,P_k\}$ be a path decomposition of $G$ such that every vertex $v\in V$ belongs to exactly one of the paths of $S_p$. Any path decomposition naturally splits the edges of $G$ into: (a) \emph{path edges} that connect consecutive vertices in the same path, (b) \emph{cross edges}  that connect vertices that belong to different paths, and (c) \emph{path transitive edges} that connect non-consecutive vertices in the same path. Given $S_p$, Algorithm PBH, draws the vertices of each path $P_i$ \emph{vertically aligned} on some $x$-coordinate depending on the order of path $P_i$.  There is one column between paths that is reserved for the bends (if any) of some cross edges.  Therefore, the total width of the resulting drawing is $2k-1$.
The $y$-coordinate of each vertex is equal to its order in a topological sorting of $G$.
Hence the height of the resulting drawing is $n-1$.  In the algorithms of~\cite{JGAA-502} path transitive edges are omitted from the final drawing.

OGDF is a self-contained C++ library of graph algorithms, in particular for (but not restricted to) automatic graph drawing.
The hierarchical drawing implementation of the Sugiyama framework in OGDF is implemented following~\cite{gansner1993technique,sander1996layout}.  
The Sugiyama framework in OGDF uses the following default choices: For the first phase of Sugiyama, it uses the $LongestPathRanking$ (a ranking module that determines the layering of the graph, i.e., the assignment of vertices into layers) which implements the well-known longest-path ranking algorithm.  Next, it performs crossing minimization by using $BarycenterHeuristic$. This module performs two-layer crossing minimization and is applied during the top-down and bottom-up traversals~\cite{DBLP:reference/crc/ChimaniGJKKM13}.  The crossing minimization is repeated 15 times, and keeps the best. Each repetition (except for the first) starts with randomly permuted nodes on each layer. Finally it computes the final coordinates  with $FastHierarchyLayout$ which  computes the final layout of the graph.  The two hierarchical drawings shown in Figure~\ref{teaser} demonstrate the significant differences in philosophy between the two frameworks.




\section{An Algorithm for Computing Compact Drawings}

We present an extension of the framework of~\cite{JGAA-502} by (a) compacting the drawing in the vertical direction, and (b) drawing the path transitive edges that were not drawn in~\cite{JGAA-502}.  This approach naturally splits the edges of $G$ into three categories, \emph{path edges}, \emph{cross edges}, and \emph{path transitive edges} that are drawn differently.  This clearly adds to the understanding of the user and allows a system to show the different categories separately without altering the user's mental map.

\subsection{Compaction} 
Let $G=(V,E)$ be a DAG with $n$ vertices and $m$ edges.  Following the framework of~\cite{ortali2018algorithms,JGAA-502} the vertices of $V$ are placed in a unique $y$-coordinate, which is specified by a topological sorting. Let $T$ be the list of vertices of $V$ in ascending order based on their $y$-coordinates. We start from the bottom and visit each vertex in $T$ in ascending order. For every vertex $v$ in this order we assign a new $y$-coordinate, $y(v)$, following a simple rule that compacts the height of the drawing: "If $v$ has no incoming edges then we set its $y(v)$ to $0$, else we set $y(v)$ equal to $a+1$, where $a$ is the $highest$ $y$-coordinate of the vertices that have edges incoming into $v$."   

\begin{figure}
\centering
\begin{subfigure}[b]{0.54\textwidth}
\centering
  \includegraphics[height=9cm ,width=0.55\linewidth]{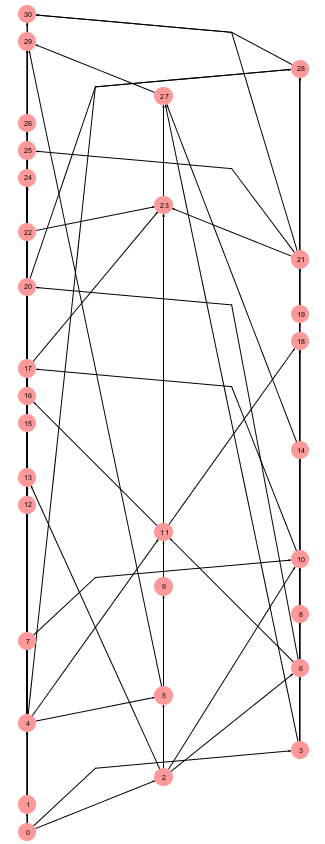}
  \caption{ }
  \label{fig:sfig1}
\end{subfigure}%

\begin{subfigure}[b]{0.55\textwidth}
\centering
  \includegraphics[height=6cm ,width=0.5\linewidth]{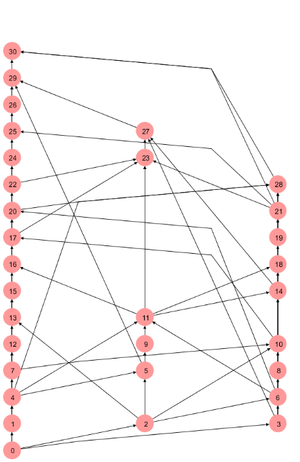}
  \caption{}
  \label{fig:sfig2}
\end{subfigure}
\caption{DAG $G$ of Figure~\ref{teaser} drawn without its path transitive edges: (a) drawing $\Gamma_1$ is computed by Algorithm PBH, and it is the input of Algorithm \ref{alg:compactionAlgo}, (b) drawing $\Gamma_2$ is the output of Algorithm \ref{alg:compactionAlgo}.}
\label{steps}
\end{figure}

Algorithm \ref{alg:compactionAlgo} takes as input a DAG $G$, and a path based hierarchical drawing $\Gamma_1$ of $G$ computed by Algorithm PBH and it produces as output a new, compacted, path based hierarchical drawing $\Gamma_2$ with height $L$, where $L$ is the length of a longest path in $G$.  Clearly this simple algorithm can be implemented in $O(n+m)$ time. Figure~\ref{steps} shows an example of two hierarchical drawings of the same graph: $\Gamma_1$ is before compaction and $\Gamma_2$ is after compaction. 

\begin{algorithm}
\label{alg:compactionAlgo}
{\bf Input:} A DAG $G=(V,E)$, and a path based hierarchical drawing $\Gamma_1$ of $G$ computed by Algorithm PBH \newline
{\bf Output:} A compacted path based hierarchical drawing $\Gamma_2$ with height $L$, where $L$ is the length of a longest path in $G$.
\label{algo}
\begin{algorithmic}[1]
    \FOR{$v \in G$:}
        \STATE Let $E_v$ be the set of incoming edges, $e=(w,v)$, into $v$:
        \IF{$E_v=\emptyset$}
        \STATE $y(v)$=0
        \ELSE
        \STATE $y(v)$=max\{$y$-coordinates of vertices $w$ with $(w,v) \in E_v$\} + 1
        \ENDIF
    \ENDFOR
\end{algorithmic}
\end{algorithm}


Notice that the first case of the if-statement, is executed only for the first vertex (source) of some paths. Clearly, the rest of the vertices have at least one incoming edge since they belong to some path where every vertex is connected to its predecessor. This is the case for the "else" part. The compacted $y$-coordinate for the rest of the vertices will always be equal to "max \{y coordinates of adjacent vertices to it\} +1". Based on these statements and the fact that the drawing after compaction is also a path  based  hierarchical drawing, we have the next two simple lemmas.

 \begin{lemma}
 \label{lemma1}
     Two vertices of the same path cannot have the same $y$-coordinate.
 \end{lemma}
 
  \begin{lemma}
  \label{lemma2}
    For every vertex $v$ with $y(v)\neq 0$, there is an incoming edge into $v$ that starts from a vertex $w$  such that $y(v) = y(w) + 1$.
 \end{lemma}

Based on these lemmas the height of the compacted drawing of the graph $G$ is at most $L$:
  \begin{theorem}
   \label{th:height}
    Let $G=(V,E)$ be a DAG with $n$ vertices and $m$ edges. Algorithm Compaction computes in $O(n+m)$ time a hierarchical drawing $\Gamma_2$ of $G$ with height $L$, where $L$ is equal to the length of a longest path in $G$.
 \end{theorem}

\begin{proof}
It is clear that the height of the resulting drawing $\Gamma_2$ cannot be lower that $L$, the length of the longest path, due to Lemma~\ref{lemma1} and the fact that all edges go from a vertex with lower to a vertex with higher $y$-coordinate. 
Similarly,  the height of the resulting drawing $\Gamma_2$ cannot be higher that $L$ since that would imply that there is a $y$ coordinate that does not contain a vertex of a longest path. In this case by the initial assumption and Lemma~\ref{lemma2} there is another path that is longer than $L$.  Hence the  height of the resulting drawing $\Gamma_2$ is equal to $L$.
The time complexity of Algorithm Compaction is immediate from the fact that we visit each vertex exactly once, in the order specified by $T$ and consider all its incoming edges once.
\end{proof}



\subsection{Drawing the Path Transitive Edges}
An important aspect of our work is the preservation of the mental map of the user that can be expressed by the reachability information of a DAG. At this point, we highlight that for every decomposition path, we have a set of path transitive edges that are not drawn by the framework of~\cite{ortali2018algorithms,JGAA-502}. In this subsection we show how to draw these edges while preserving the user's mental map of the previous drawing.  Additionally, one may interact with the drawings by hiding the path transitive edges at the click of a button without changing the user's mental map of the complete drawing.

Now we will describe an algorithm that draws the path transitive edges using the minimum extra width (minimum extra number of columns) for each decomposition path. The steps of the algorithm are briefly described as follows:

\begin{enumerate}
\item For every vertex of each decomposition path we calculate the indegree and outdegree based only on path transitive edges, i.e., excluding path edges and cross edges.
\item  If all indegrees and outdegrees are zero the algorithm is over, if not, we select a vertex $v$ with the highest indegree or outdegree and we bundle all the incoming or outgoing edges of $v$, respectively. These bundled edges are represented by an $interval$ with starting and finishing points, the lowest and highest $y$-coordinates of the vertices, respectively.
\item Next, we insert each interval on the left side of the path on the first available column such that the interval does not overlap with another interval (see details below).
\item   We remove these edges from the set of path transitive edges, update the indegree and outdegree of the vertices and repeat the selection process.
\item The intervals of the rightmost path, are inserted on the right side of the path in order to avoid potential crossing with cross edges. 
\item A final, post-processing step can be applied because some crossings between intervals/bundled edges can be removed by changing the order of the columns containing them.
\end{enumerate}

\begin{figure}[htp]
\begin{subfigure}{.2\textwidth}
  \centering
  \includegraphics[scale=0.2]{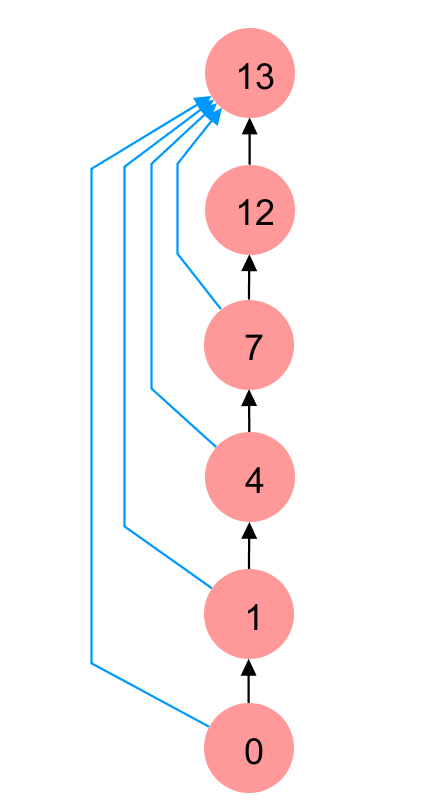}
  \caption{}
  \label{fig0:sfig1}
\end{subfigure}
\begin{subfigure}{.21\textwidth}
  \centering
  \includegraphics[scale=0.2]{{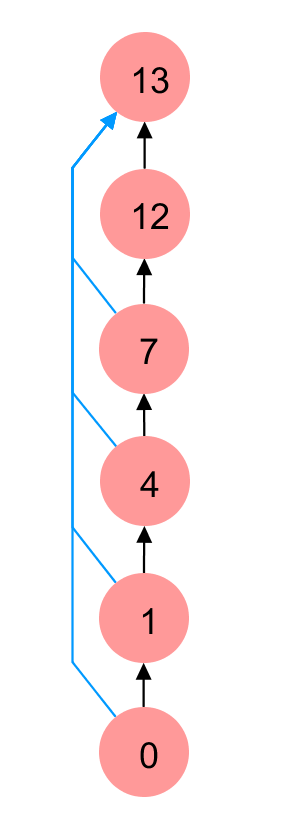}}
  \caption{}
  \label{fig1:sfig2}
\end{subfigure}
\hfill
\begin{subfigure}{.2\textwidth}
  \centering
  \includegraphics[scale=0.2]{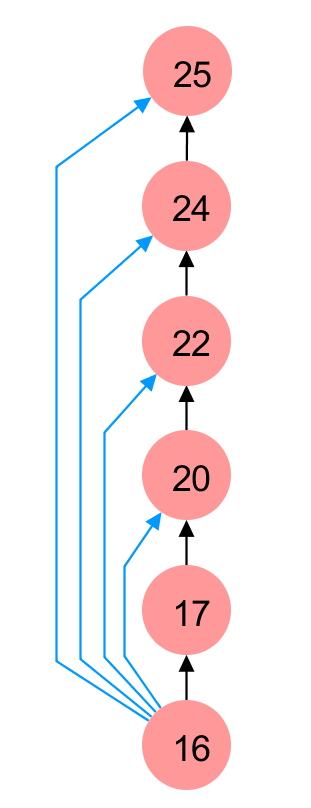}
  \caption{}
  \label{fig0:sfig1}
\end{subfigure}
\begin{subfigure}{.35\textwidth}
  \centering
  \includegraphics[scale=0.2]{{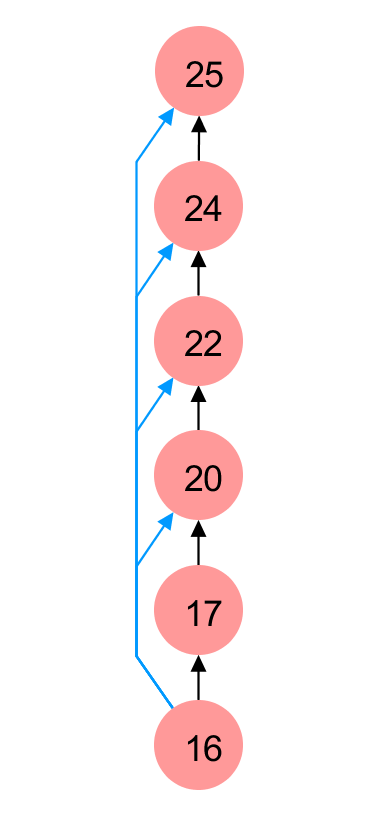}}
  \caption{}
  \label{fig1:sfig2}
\end{subfigure}
\caption{Bundling of path transitive edges: (a) incoming edges into node 13, (b) after bundling, (c) outgoing edges from node 16, (d) after bundling. }
\end{figure}


The above algorithm can be implemented to run in time $O(m+n \log n)$ by handling the updates of the indegrees and outdegrees carefully, and placing the appropriate intervals in a (Max Heap) Priority Queue.
As expected, the fact that we draw the path transitive edges increases the number of bends, crossings, and area, with respect to not drawing them.


For each decomposition path, suppose we have a set of $b$ intervals such that each interval $I$ has a start point, $s_I$, and a finish point $f_I$. The starting point is the position of the vertex of the interval with the lowest $y$-coordinate. Similarly, the finish point is the position of the node of the interval with the highest $y$-coordinate.  We follow a greedy approach in order to minimize the width (number of columns) for placing the bundled edges.  The approach is similar to Task Scheduling~\cite{GTbook}, for placing the intervals. It uses the optimum number of columns and runs in $O(b \log b)$ time, for each path with $b$ intervals. This is done by considering the intervals of each decomposition path in increasing order of their starting points. We select each interval (resp. task) according to its starting point and place it into the first column that can fit (i.e., does not intersect with another interval). If there are no available columns, we allocate a new column and place the interval there.  Since the sum of all $b$'s for all paths in a path decomposition is at most $n$ we conclude that the algorithm runs in $O(n \log n)$ time.  The proof of correctness is similar to the one for Task Scheduling in~\cite{GTbook} and thus it is omitted here.

\begin{theorem}
   \label{th:width}
    Let $G=(V,E)$ be a DAG with $n$ vertices and $m$ edges. There is an algorithm that computes a drawing of $G$ bundling the path transitive edges for each path using the minimum number of columns (width) per path. The algorithm runs in $O(m + n \log n)$ time and computes a compact hierarchical drawing of $G$.
 \end{theorem}

\section{Experimental Results and Comparisons}


We performed experiments in order to compare the results produced by the two frameworks on different DAGs with varying number of nodes and edges. We use 20 DAGs that were produced in a random, but controlled, fashion in order to have small and large DAGs, but with a predefined  average degree. Furthermore, in order to evaluate the performance of the two drawing frameworks, we use the following standard metrics:

\begin{description}
\item[$\bullet$ Number of crossings.]
\item[$\bullet$ Number of bends.]
\item[$\bullet$ Width of the drawing:] The total number of distinct x coordinates that are used by the framework.
\item[$\bullet$ Height of the drawing:]The total number of distinct y coordinates that are used by the framework.  
\item[$\bullet$ Area of the drawing:] The area of the enclosing rectangle.
\end{description}


\begin{figure}[htp]
\includegraphics[width=1\linewidth,height=9.5cm]{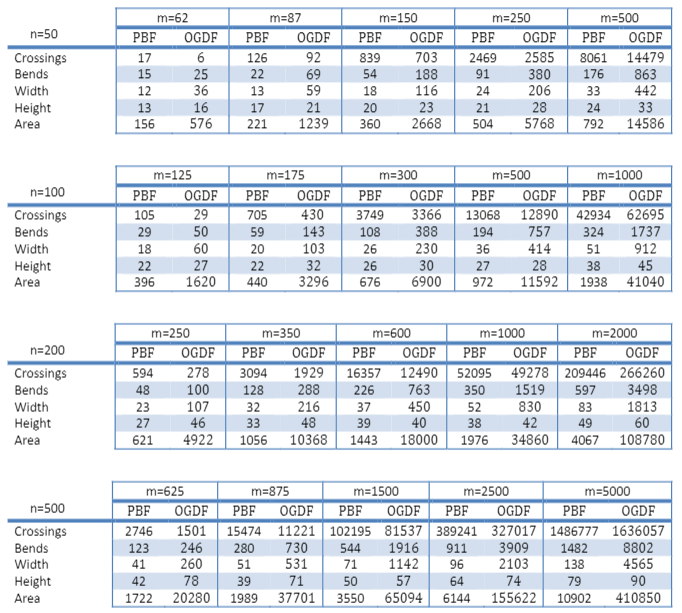}
\caption{Results on \textit{number of crossings, bends, width, height} and \textit{area} for \textit{PBF} and \textit{OGDF} for all DAGs in our study.}
\label{tablePDF}
\end{figure}

\begin{figure}[htp]
\includegraphics[width=1\linewidth,height=7cm]{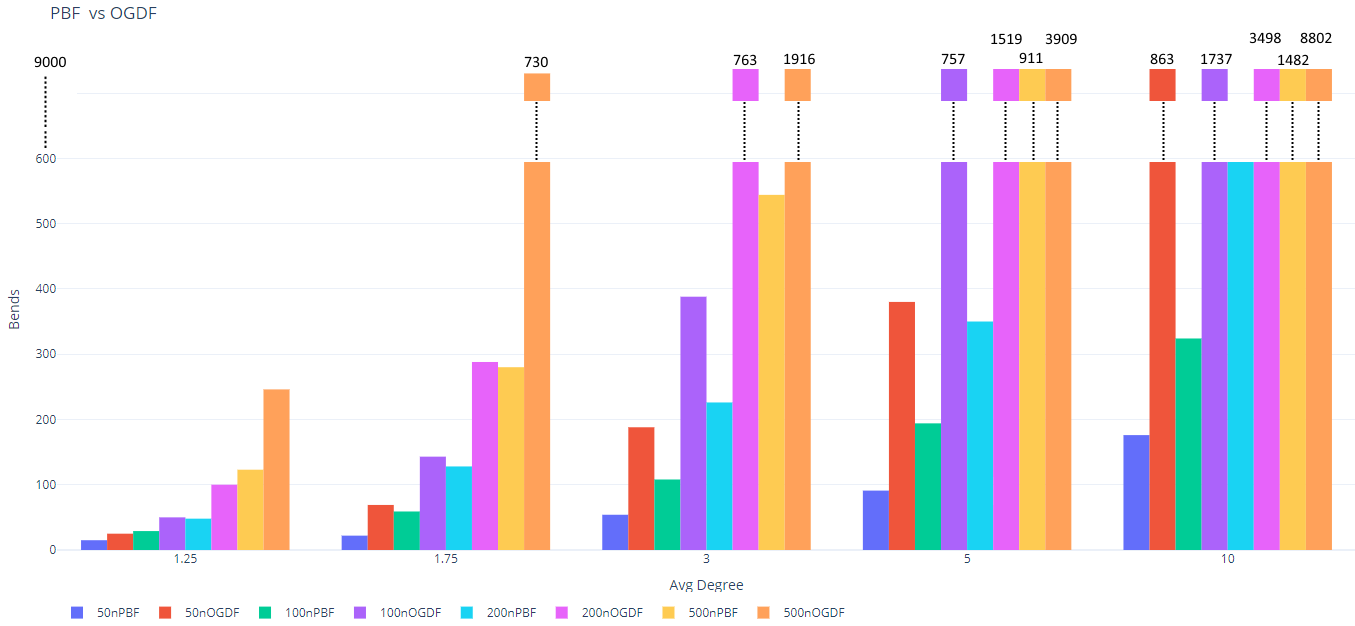}
\caption{Results on the \textit{number of bends for \textit{PBF} and \textit{OGDF} for all DAGs in our study.}}
\label{bends_all}
\end{figure}

Figure~\ref{tablePDF} shows a table that contains the results of our experiments based on these metrics for $PBF$ as implemented in TS Perspectives~\cite{Tom} compared to the results produced by OGDF. In order to be consistent  with  the  experimental  settings of OGDF, we used the default parameters.  In the experiments that we present in this section we see that in all cases our approach gives better results than the ones produced by OGDF with respect to the number of bends, width, height, and as expected the total area of the drawings. For the number of bends we observe that our proposed technique produces bends that are a small fraction of $n$, whereas OGDF produces bends that are proportional to $m$. The bar charts shown in Figure~\ref{bends_all} show how the number of bends grows as the DAGs grow in size and average degree and provide a clear evidence that the number of bends for $PBF$ is significantly lower than OGDF in all cases.  On the other hand, the drawings of OGDF have a lower number of crossings when the input graphs are relatively sparse.  However, when the graphs are a bit denser (e.g., average degree higher than five) our drawings start having less crossings.
Since the two frameworks use a different coordinate system, for a fair comparison between them we chose to count as height of a drawing the number of different layers (or different $y$-coordinates) and as width the number of different $x$-coordinates of nodes and bends, used by each system.  In other words, we normalize the two coordinate systems by mapping them on a "grid." 

In general, our experiments show that $PBF$ produces readable drawings with very good results almost in all metrics, except for the number of crossings. 
Additionally, it clearly partitions the edges into three distinct categories, and vertically aligns certain paths, which can be user defined. This can be a great advantage in certain applications and therefore it seems to be an interesting alternative, as also shown in Figure 6 for a larger example. 
$PBF$ does not perform any crossing reduction step, in contrast to OGDF which offers crossing minimization algorithms by default (also required by the Sugiyama framework), which are run several times in order to keep the best result. 

\subsection{A Heuristic for Ordering the Paths:}
As described in~\cite{ortali2018algorithms}, one way to minimize the number of crossings between cross edges and path edges (and path transitive edges, now) is to check all possible $k!$ permutations of the $k$ paths.
In order to reduce the number of crossings, we implemented a heuristic that aims to reduce the number of paths crossed by cross edges.  Our fast and simple approach is described below.

We create an undirected \emph{path graph} by placing a node for each path $P$.  For any pair of paths $P_1$ and $P_2$ we find the total number of cross edges between them, $c$, and we insert an (undirected) edge between the nodes corresponding to paths $P_1$ and $P_2$ with weight equal to $c$. Hence, the weight, $c$, of edge $(P_1,P_2)$ is the sum of the number of cross edges from $P_1$ to $P_2$ plus the number of cross edges from $P_2$ to $P_1$.
We do this for all cross edges between all paths.
Next, we order the paths following a greedy process:  We find the maximum-weight edge and we place the corresponding paths next to each other. We remove the edge from the path graph and continue with this process until it contains no edges. If we select an edge such that both paths are already placed, we simply delete this edge and proceed. If we select an edge such that one of the two paths is not already placed, then we place it at the rightmost (or leftmost) side of the placed path, depending upon which side has the least number of paths placed.  This algorithm uses data structures similar to Kruskal's~\cite{kruskal1956shortest} algorithm for computing a minimum (maximum) spanning tree and it can be implemented in $O(m + k \log k)$ time.
We performed some limited experiments on sparse graphs (with average degree 1.25, 1.75, and 3) using this path ordering algorithm, and we found out that the produced drawings have lower number of crossings, bends, and edge length.  Unfortunately, for denser graphs the results are inconclusive.

\section{Conclusions and Open Problems}
\label{Section:Conclusion}
We present algorithms and experimental results comparing two hierarchical drawing frameworks: (a) the path-based framework and (b) OGDF, which is based on the Sugiyama technique.  
Our compaction algorithm runs in $O(km)$ time, and produces drawings with height equal to the length of a longest path of $G$ instead of $n-1$ which is the height of drawings produced in~\cite{JGAA-502}.
In this implementation we present an algorithm to bundle and draw the path transitive edges of $G$ in $O(m + n \log n)$ time, which is an extension of the original path based framework~\cite{JGAA-502}.
The experimental results show that the drawings produced by our algorithms have significantly lower number of bends and are much smaller in area than the ones produced by OGDF, but they have more crossings for sparse graphs.  Thus our algorithms offer an interesting alternative when we visualize hierarchical graphs. They focus on showing important aspects of the graph such as critical paths, path transitive edges, and cross edges. For this reason, this framework is particularly useful in graph visualization systems that encourage user interaction.
There are several interesting open problems:
1) Find better algorithms to order the paths. 2) Find techniques to reduce the number of crossings.  3) Allow some extra vertical space between selected vertices in order to make the visualization more visually appealing.

\clearpage
\bibliographystyle{splncs04}
\bibliography{main} 
\end{document}